\documentclass[twocolumn,amsmath,amssymb,floatfix,prb,showpacs,footinbib,superscriptaddress]{revtex4}
\usepackage{amsmath,amssymb,natbib,bm,graphicx,url,psfrag,times}
\usepackage{subdepth}
\usepackage[hyperfigures,colorlinks=true,allcolors=blue]{hyperref}

\newcommand{\RR}{\mathbb{R}}

\newcommand{\abs}[1]{\left|#1\right|}
\newcommand{\bra}[1]{\langle \, #1 \,|}
\newcommand{\ket}[1]{|\, #1 \, \rangle}
\newcommand{\bket}[2]{\langle \, #1 \,|\, #2 \, \rangle}
\newcommand{\boket}[3]{\langle\, #1 \,|\, #2 \,|\, #3 \,\rangle}

\newcommand{\be}{\begin{equation}}
\newcommand{\ee}{\end{equation}}
\newcommand{\bc}{\begin{center}}
\newcommand{\ec}{\end{center}}

\begin{document}
\title{Asymptotic Expressions for Charge Matrix Elements of the Fluxonium Circuit}
\author{Guanyu Zhu}
\affiliation{Department of Physics and Astronomy, Northwestern University, Evanston, Illinois 60208, USA}
\author{Jens Koch}
\affiliation{Department of Physics and Astronomy, Northwestern University, Evanston, Illinois 60208, USA}
\date{\today}

\begin{abstract}
In charge-coupled circuit QED systems, transition amplitudes and dispersive shifts are governed by the matrix elements of the charge operator. For the fluxonium circuit, these matrix elements are not limited to nearest-neighbor energy levels and  are conveniently tunable by magnetic flux. Previously, their values were largely obtained numerically. Here, we present analytical expressions for the fluxonium charge matrix elements. We show that new selection rules emerge in the asymptotic limit of large Josephson energy and small inductive energy. We illustrate the usefulness of our expressions for the qualitative understanding of charge matrix elements in the parameter regime probed by previous experiments.
\end{abstract}
\maketitle

\section{Introduction}
The fluxonium circuit \cite{manucharyan_fluxonium:_2009,phase_slip_2012} is one of the most recent additions to the family of superconducting qubits. It is composed of a small Josephson junction shunted by a large Josephson junction array which primarily acts as a large kinetic inductance.  For quantum state manipulation and readout, the fluxonium circuit can be capacitively coupled to a microwave resonator and thus integrated into the circuit QED architecture \cite{blais_cavity_2004,schoelkopf_wiring_2008}. The amplitudes of photon-induced transitions between different energy levels are then determined by the charge matrix elements $\mathsf{N}_{ll'}=\boket{l}{ \mathsf{N}}{l'}$ where $l$ and $l'$ denote the circuit's eigenstates, and $\mathsf{N}$ is the dimensionless charge operator. For circuits like the Cooper pair box (CPB) in both charging \cite{nakamura_coherent_1999,bouchiat_quantum_1998} and transmon regime \cite{koch_charge-insensitive_2007,schreier_suppressing_2008}, simple selection rules give a very good approximation limiting the one-photon transitions to nearest-neighbor levels ($l\!\to\!l\pm1$). For the fluxonium circuit, this selection rule is absent  --  leading to interesting and useful features including the experimentally observed large dispersive shifts over a wide external flux range despite strong detuning between the lowest fluxonium energy splitting and the photon frequency. \cite{phase_slip_2012,guanyu_dispersive2012}.

In previous work, we have presented numerical results for the charge matrix elements of the fluxonium circuit  \cite{koch_charging_2009,guanyu_dispersive2012}. As illustrated in Ref.~\onlinecite{guanyu_dispersive2012} with results obtained for the experimentally realized parameter values of Josephson, charging and inductive energy, matrix elements indeed do not obey strict selection rules. Nonetheless, trends of  certain matrix elements being up to an order of magnitude larger than others hint at the fact that a new set of selection rules emerges asymptotically in the limit of large Josephson energy and small inductive energy.
In this limit, and making use of the classification of fluxonium eigenstates into metaplasmon and persistent-current states \cite{koch_charging_2009}, we derive analytical expressions for the charge matrix elements. Based on the asymptotic selection rules, we finally shed light on the different magnitudes of charge matrix elements  realized in the experimental parameter regime.

Our paper is structured as follows. In Sec.~\ref{analytical},  we briefly summarize the classification of the fluxonium eigenstates into metaplasmon and persistent-current states (previously presented in Ref.\ \onlinecite{koch_charging_2009}) and derive analytical expressions for the charge matrix elements. Based on the resulting asymptotic selection rules, we distinguish matrix elements of different magnitudes and compare the analytical results with numerical results for the experimentally realized parameters in  Sec.~\ref{application}. Finally, we summarize our findings in Sec.~\ref{sec:conclusion}.

\section{Analytical expressions for fluxonium charge matrix elements}\label{analytical}
The Hamiltonian describing the fluxonium circuit within the superinductance model \cite{koch_charging_2009, ferguson_2012} is given by
\be\label{Hf}
H_\text{f} = 4E_C  {\mathsf{N}}^2 -E_J \cos { {\varphi}}+\frac{1}{2}E_L( {\varphi}+2\pi\Phi_{\text{ext}}/\Phi_0)^2.
\ee
Here,  the operator $ {\varphi}$ describes the phase difference across the small junction. The conjugate operator $ \mathsf{N}=-i\frac{d}{d\varphi}$ is associated with the charge imbalance across the small junction, in units of the Cooper pair charge $(2e)$.   The three coefficients represent the three relevant energy scales in the circuit, namely the charging energy $E_C=e^2/(2C_J)$, Josephson energy $E_J$ of the small junction, and the effective inductive energy $E_L$ of the ``superinductor" made by the Josephson junction array. 
It is instructive to view the Hamiltonian $H_\text{f}$ as describing a fictitious particle in a sinusoidal potential, deformed by an overall parabolic envelope. In this point of view, $\varphi$ plays the role of the spatial coordinate. Hence, the Josephson and  inductive energy terms in $H_\text{f}$ determine the potential energy $V(\varphi)$, while the charging term produces the kinetic energy contribution. The external magnetic flux $\Phi_\text{ext}$ (in units of the flux quanta $\Phi_0\!=\!h/2e$) spatially shifts the parabolic envelope. Due to the presence of the inductive term, the appropriate boundary conditions supplementing the stationary Schr\"odinger equation for $H_\text{f}$ are derived from normalizability of its eigenstates $\ket{\psi}$, i.e., $\int_\RR d\varphi\,|\bket{\varphi}{\psi}|^2<\infty$, implying $\psi(\varphi) \rightarrow 0$ when $\varphi \rightarrow \pm \infty$.

In the limit of large Josephson and small inductive energy,  
\be\label{condition}
 E_J  \gg E_C \gg E_L,
 \ee
the low-lying eigenstates of fluxonium can be classified into two physically distinct types:  \emph{metaplasmon states} and \emph{persistent-current states} \cite{koch_charging_2009}.
For clarity and introduction of necessary notation, we briefly review this classification as obtained in Ref.\ \onlinecite{koch_charging_2009}. To do so, we rewrite $H_\text{f}$ in a more suitable basis and start by separating off the inductive energy term, $H_\text{f}=H'+H_\text{ind}$, where $H'= 4E_C  {\mathsf{N}}^2 -E_J \cos { {\varphi}}$.  Despite the tempting appearance of $H'$, we must refrain from identifying it as the ordinary Cooper pair box Hamiltonian: in Eq.\ \eqref{Hf}, the spatial coordinates $\varphi$ and $\varphi+2\pi$ are distinct positions. Hence, $H'$ is not subject to periodic boundary conditions as the Cooper pair box, but rather obeys the quasi-periodic boundary conditions familiar from Bloch's theorem, as appropriate for a particle in an infinitely extended periodic potential. Accordingly, the eigenstates of $H'$ are Bloch states,
\be
H' \ket{p,s} = \epsilon_s(p) \ket{p,s},
\ee 
where $s \in \mathbb{N}$ is the band index, $p \in [0,1)$  the quasimomentum in the first Brillouin zone, and $\epsilon_s(p)$ denotes the band dispersion for the cosine potential (which coincides with the ordinary offset charge dispersion of the Cooper pair box levels \cite{koch_charge-insensitive_2007}).

To rewrite the inductive contribution $H_\text{ind}$ in the Bloch basis, we re-interpret $p$ as a new spatial coordinate. Since it ``lives'' on a circle with circumference $1$, the resulting expression $\varphi=i\,d/dp+\Omega$ for the phase operator must generally include an inter-band coupling operator $\Omega$ \cite{lifshitz_statistical_1980}. This inter-band coupling can be neglected for low-lying bands and sufficiently large $E_J/E_C$ \cite{koch_charging_2009}. In that limit, the Hamiltonian $H_\text{f}$, hence, becomes block-diagonal,  splitting into individual Hamiltonians for each band $s$, $H_\text{f}\approx \sum_s H_s \ket{s}\bra{s}$, where
\be\label{Hs1}
H_s = \frac{E_L}{2} \left(i \frac{d}{dp} + \frac{2\pi\Phi_\text{ext}}{\Phi_0}\right)^2 + \epsilon_s(p).
\ee
Now, accompanied by periodic boundary conditions in $p$,  each Hamiltonian $H_s$ indeed has the same structure as the Hamiltonian of a Cooper pair box. The only difference lies in the form of the periodic ``potential'' $\epsilon_s(p)$, which generally deviates from a pure cosine. To make the analogy concrete, note that  the variable $2\pi p \in [0,2\pi)$ in $H_s$ takes on the role of  the periodic phase variable of the Cooper pair box, and the external flux $\Phi_\text{ext}/\Phi_0$ that of the Cooper pair box offset charge $n_g$.

Next, two different types of low-lying fluxonium states can be distinguished for each band $s$. First, eigenstates $\ket{\nu,s}$ with energies  below the maximum of the energy dispersion, $E_{\nu s}(\Phi_\text{ext}) <\text{max}_p\epsilon_s(p) $, are \emph{metaplasmon states}.  They are quasi-bound states \footnote{We speak of  ``quasi-bound'' states to acknowledge that boundedness is not perfectly well-defined for wavefunctions on a closed space.} in the $\epsilon_s(p)$ potential analogous to the lowest states of the Cooper pair box in the transmon regime. The corresponding eigenenergies depend only weakly on the external flux $\Phi_\text{ext}$, just as Cooper pair box levels are offset-charge insensitive in the transmon regime \cite{koch_charge-insensitive_2007}.  Second,  eigenstates 
with energies above the maximum of the energy dispersion,
are \emph{persistent-current states}. Their energies strongly depend on the external flux $\Phi_\text{ext}$, closely mimicking the offset-charge dependence of the high-lying transmon levels (for which, effectively, the charging regime holds). While quasi-itinerant in the $\epsilon_s(p)$ potential, persistent-current states localize in the individual minima of the $V(\varphi)$ potential [Eq.\ \eqref{Hf}]. They are conveniently expressed in terms of  Wannier states
\be
\ket{m,s} = \int_{-\frac{1}{2}}^{\frac{1}{2}} dp\, e^{- i 2\pi m p} \ket{p,s}.
\ee
Expressed in this basis, the Hamiltonian Eq. \eqref{Hs1}  reads:
\begin{align}\label{Hs2}
 H_s \approx &\frac{(2\pi)^2}{2} E_L ( \mathsf{m}+\Phi/\Phi_0)^2 \\\nonumber
& + \frac{1}{2}\sum_{m=-\infty}^{\infty}\epsilon_{s,1}\bigg[\ket{m,s}\bra{m+1,s}+\text{H.c.}\bigg]+\epsilon_{s,0},
\end{align}
where we have  approximated the potential $\epsilon_s(p)$ by the truncated Fourier series
$
\epsilon_s (p) \approx \epsilon_{s,0} + \epsilon_{s,1}\cos({2\pi p}),
$
and used  $\sum_{m=-\infty}^{\infty}\ket{m,s}\bra{m\!\!+\!\!1,s}=e^{-i2\pi p}$ as well as  $\mathsf{m}=i\,d/d(2\pi p)$. Note that in the transmon regime ($E_J \gg E_C$), $\epsilon_{0,0}$ is just the plasmon energy $\sqrt{8E_J E_C}$. Analytical approximations for $\epsilon_{s,0}$ and $\epsilon_{s,1}$ in the transmon regime are given in Ref.\ \onlinecite{koch_charge-insensitive_2007}.
Based on the classification into metaplasmon and persistent-current state, we are ready to derive analytical expressions and asymptotic selection rules for the charge matrix elements. Due to the two types of states involved, there are three possible types of charge matrix elements, which we discuss one by one in the following.

\paragraph{Matrix elements between persistent-current states.} 
The Wannier states $\ket{m,s}$ provide good approximations for the persistent-current states (away from degeneracies which occur at integer and half-integer $\Phi_\text{ext}/\Phi_0$).
The charge matrix elements between two persistent-current states, possibly from different bands $s$ and $s'$, are then given by
\begin{align}\label{pp1}
&\boket{m,s}{ \mathsf{N}}{m',s'} \approx -i\left(\frac{E_J}{32E_C}\right)^{\frac{1}{4}}\int_{-\infty}^{\infty}d\varphi \, w_{ms}^*(\varphi)\\\nonumber
&\qquad\qquad\times \bigg[\sqrt{s'}\,w_{m' s'-1}(\varphi)-\sqrt{s'+1}\,w_{m' s'+1}(\varphi)\bigg].
\end{align}
Here, $w_{ms}(\varphi) \equiv \bket{\varphi}{m,s}$ is the approximate persistent-current state wavefunction in $\varphi$-space. Due to the  strong localization in the minima of $V(\varphi)$, persistent-current states in adjacent minima are nearly orthogonal. One, hence, obtains the approximation
\begin{align}\label{pp2}
& \boket{m,s}{ \mathsf{N}}{m',s'} \\\nonumber
&\quad  \approx   -i\left(\frac{E_J}{32E_C}\right)^{\frac{1}{4}}\delta_{m,m'} \bigg[\sqrt{s'}\delta_{s,s'-1}-\sqrt{s}\,\delta_{s,s'+1}\bigg].
\end{align}
  To obtain nonzero values for the charge matrix elements in this limit, the two states involved must obey two asymptotic selection rules. The first is the neighboring-band selection rule, demanding $\Delta s= s'-s=\pm 1$. The second is the same-minimum selection rule, given by $\Delta m=m'-m=0$. Accordingly, both states involved must belong to the same local minimum $m$ of the potential $V(\varphi)$. This rule implies that the circulating persistent current (and the flux it generates) cannot change its magnitude or direction during the transition. Both rules follow intuitively from considering the momentum matrix elements of local harmonic oscillators with negligible neighbor overlap.

\paragraph{Matrix elements between metaplasmon states.} The charge matrix elements involving metaplasmon states only, can be brought into the form
\begin{align}
\nonumber \boket{\nu',s'}{ \mathsf{N}}{\nu,s} \approx \, &i\left(\frac{E_J}{32E_c}\right)^{\frac{1}{4}}(\sqrt{s}\delta_{s,s'+1}-\sqrt{s'}\delta_{s,s'-1})\\
& \times \int_{-\frac{1}{2}}^{\frac{1}{2}}dp \ \chi_{\nu's'}^*(p)\,\chi_{\nu s}(p).
\label{mm}
\end{align}
This expression was  previously derived in Ref.\ \onlinecite{koch_charging_2009}, except for a misprint in the prefactor (fixed here). By $\chi_{\nu s}(p) \equiv \bket{p,s}{\nu,s}$, we denote the metaplasmon wavefunctions in the Bloch basis. The index $\nu=0,1,2,\ldots$ labels energy levels within a fixed band $s$.  The first asymptotic selection rule manifest in Eq.\ \eqref{mm}, is the neighboring-band selection rule $\Delta s=\pm1$.
The matrix elements still depend on the overlap between two metaplasmon states, see again Ref.\ \onlinecite{koch_charging_2009} for analytic approximations and asymptotic selection rules in $\nu,\nu'$.

\paragraph{Matrix elements between metaplasmon and persistent-current states.} The last type of matrix elements involves both a metaplasmon and a persistent-current state. Its asymptotic expression is given by
\begin{align}\label{pm}
& \boket{\nu,s}{ \mathsf{N}}{m,s'} \\
\nonumber  &\approx  - \frac{i^{\nu+1}}{\sqrt{2^{\nu}\nu!}}\left(\frac{E_J E_L}{32E_C\abs{\epsilon_{s,1}}}\right)^{\frac{1}{4}}(\sqrt{s}\delta_{s,s'+1}-\sqrt{s'}\delta_{s,s'-1}) \\
\nonumber &\qquad\times \exp \left[ -\pi F_{ms}^2(\Phi_\text{ext})\right]
H_{\nu}\left[\sqrt{2\pi} F_{ms}(\Phi_\text{ext})\right].
\end{align}
Here, $H_{\nu}(x)$ is the Hermite polynomial of order $\nu$ and we have abbreviated $F_{ms}(\Phi_\text{ext})=(m+\Phi_\text{ext}/\Phi_0) ( E_L / |\epsilon_{s,1}|)^{1/4}$. The only selection rule present is the one for neighboring bands. The magnitude of the matrix elements depends on both  quantum numbers $m$ and $\nu$, and is conveniently tunable by magnetic  flux $\Phi_\text{ext}$. 

\section{Matrix elements realized in experiments}\label{application}
The values of Josephson, charging and inductive energy realized in recent experiments ($E_J/h\!\!=\!\!8.9\,\text{GHz}; E_C/h\!\! = \!\!2.48\,\text{GHz}; E_L/h \!\!= \!\!0.53\,\text{GHz}$ in Ref.\ \onlinecite{manucharyan_fluxonium:_2009}) do not quite reach the asymptotic behavior predicted for $E_J\gg E_C \gg E_L$. Hence, we cannot expect the asymptotic results from Sec.~\ref{analytical} to quantitatively match the exact results. Nonetheless, the asymptotic selection rules can still give valuable intuition and qualitative predictions for the different magnitudes of matrix elements, which will be of immediate use in the design of future fluxonium devices.

\begin{figure}
\includegraphics[width=1\columnwidth]{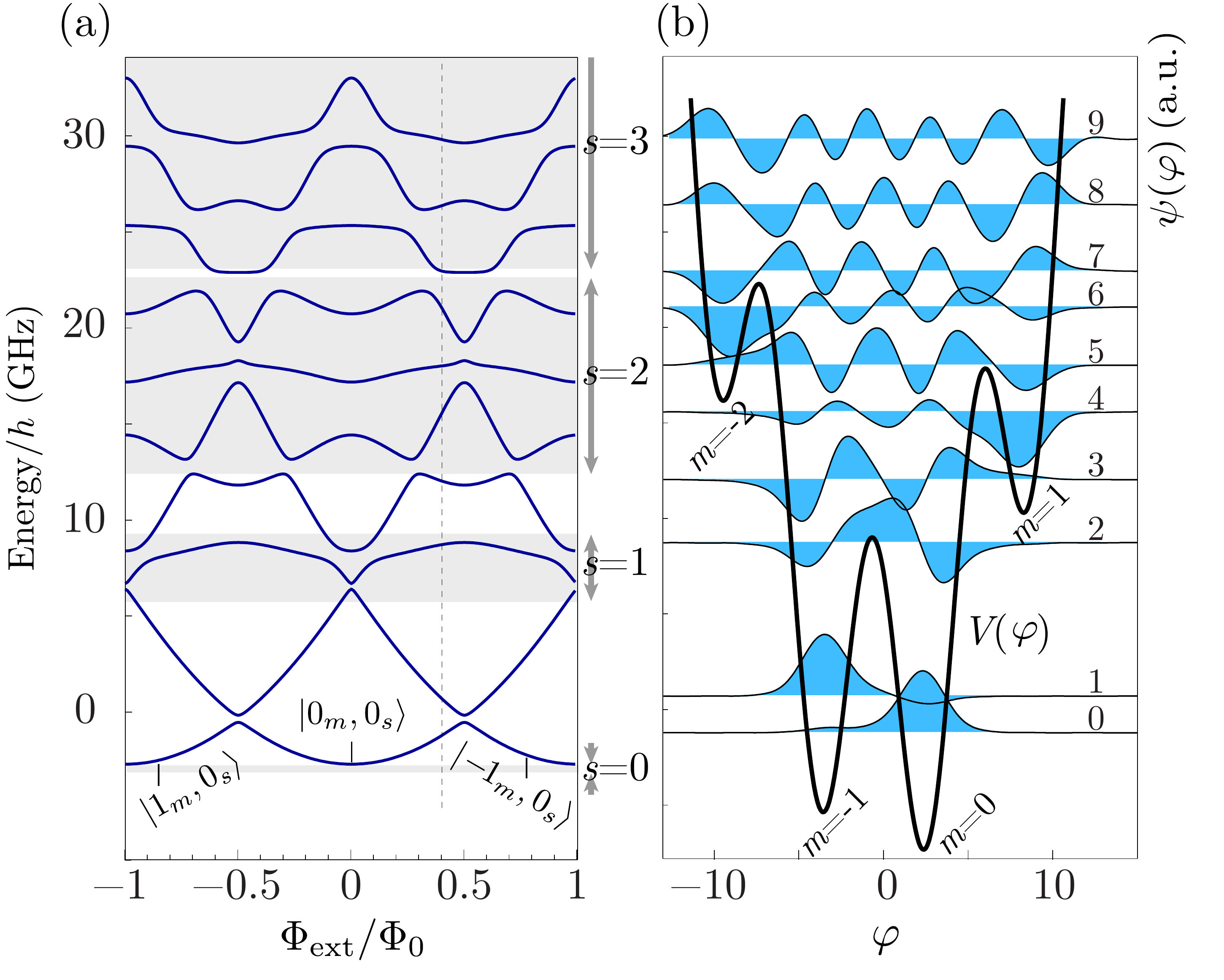}
\caption{(Color online) (a) Fluxonium energy levels (solid curves) as a function of external flux $\Phi_{\text{ext}}$, for the parameters  realized experimentally \cite{manucharyan_fluxonium:_2009}.   Shaded regions in the background show position and width of the bands $\epsilon_s(p)$. The three $s\!\!=\!\!0$ persistent-current states with parabolic flux dependence are labeled by their quantum number $m$.  (b) Fluxonium eigenfunctions for $\Phi_\text{ext}/\Phi_0\!\!=\!\!0.4$ [vertical dashed line in (a)]. The bold black curve shows the fluxonium potential $V(\varphi)$. Local minima are labeled by the quantum numbers $m$. The fluxonium wavefunctions (thin curves) are offset by their eigenenergies.
}
\label{bandmap}
\end{figure}

\begin{figure}
\centering
\includegraphics[width=0.8\columnwidth]{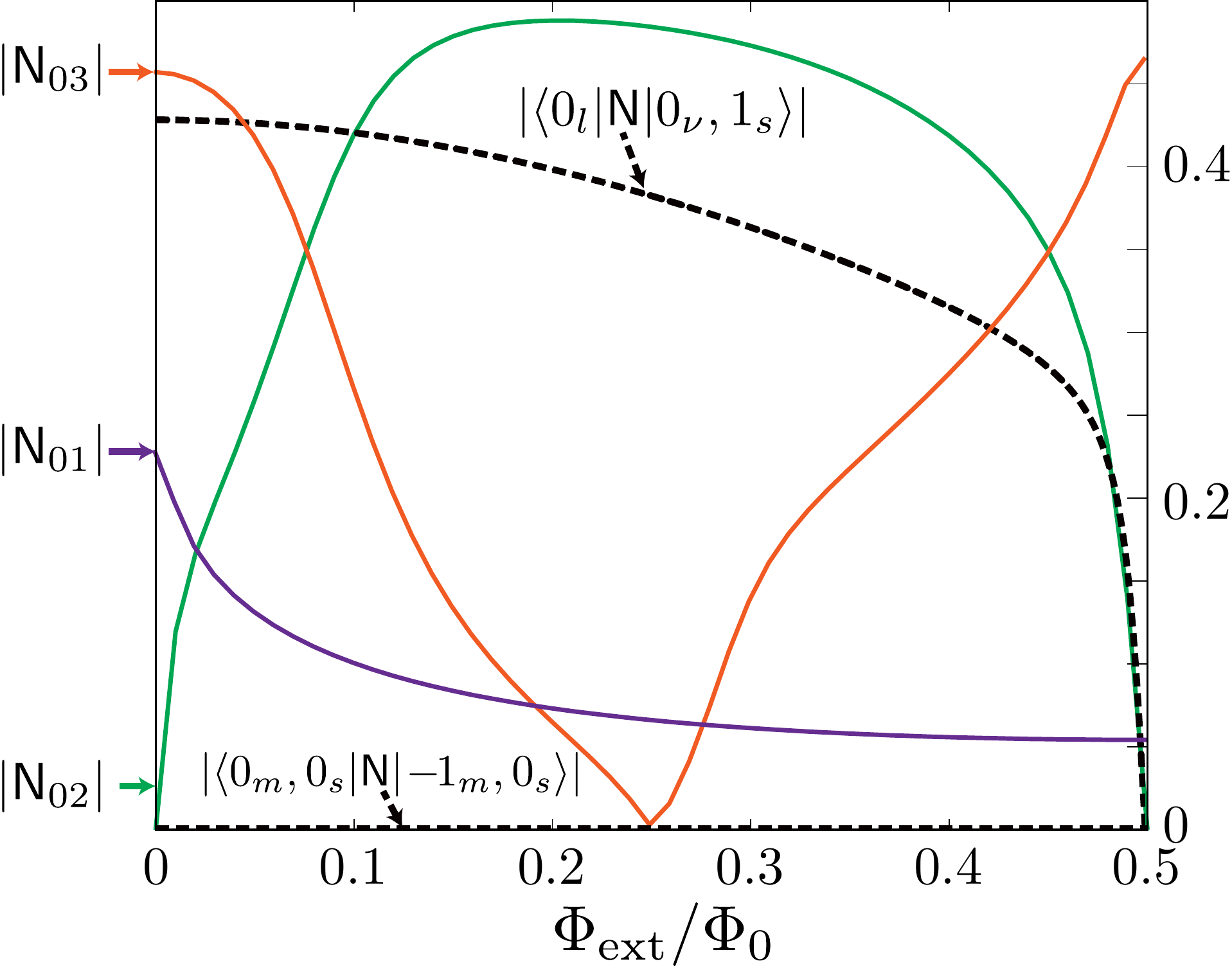}
\caption{(Color online) Comparison of numerical results and asymptotic predictions  for charge matrix elements. Solid curves show numerical results for the magnitude of the charge matrix elements, $|\langle 0_l|\mathsf{N}|1_l\rangle|$, $|\langle 0_l|\mathsf{N}|2_l\rangle|$ and $|\langle 0_l| \mathsf{N}| 3_l\rangle|$. Dashed curves show the asymptotic  matrix elements between the ground state and the lowest metaplasmon state, namely $|\langle 0_l | \mathsf{N}|0_\nu,1_s\rangle|$, and between the two low-lying persistent current states, $|\langle 0_m,0_s|\mathsf{N}|\!-\!\!1_m,0_s\rangle|$.}
\label{matrixtheory}
\end{figure}

To apply the results derived in Sec.~\ref{analytical}, we first need to establish the type  of each low-lying fluxonium eigenstate (metaplasmon versus  persistent-current), given the experimental parameters. As shown in Fig.\ \ref{bandmap}(a),  the energy dispersion of the lowest band, $\epsilon_{s=0}(p)$, turns out to be too shallow to support any metaplasmon states. As a result, the ground state and first excited state are found to be $s\!\!=\!\!0$ persistent-current states, lying in the gap between the lowest two bands $\epsilon_{s=0}(p)$ and $\epsilon_{s=1}(p)$.  
Due to inversion symmetry and periodicity in the magnetic flux, we may restrict our discussion in the following to the flux range $0\le\Phi_{\text{ext}}/\Phi_0\le 0.5$ without loss of generality. Under these conditions and sufficiently away from integer and half-integer flux, the two lowest persistent-current states are well approximated by the Wannier states            
 $|\!-\!\!1_m,0_s\rangle$ and $|0_m,0_s\rangle$. 

In the following, we focus on the example flux point $\Phi_{\text{ext}}/\Phi_0\!\!=\!\!0.4$. The exact wavefunctions at this point are illustrated in Fig.\ \ref{bandmap}(b). Note that the ground state (first excited state) is indeed primarily localized in the minimum $m\!\!=\!\!0$ ($m\!\!=\!\!-1$). 
The second and third excited states are metaplasmon states with band indices $s=1$ and $s=2$, respectively. As expected, they delocalize over multiple minima of the potential $V(\varphi)$. At $\Phi_{\text{ext}}/\Phi_0\!\!=\!\!0.4$, the fourth excited state can easily be identified as a persistent-current state of the $s\!\!=\!\!0$ band, by noting its quadratic flux dependence expected for the  $|1_m,0_s\rangle$ state.   Accordingly, it is strongly localized in the $m=1$ minimum. However, due to the large inter-band coupling for high-lying levels, this state is already significantly influenced by the nearby metaplasmon state [see the large avoided crossing of the third and fourth excited states near $\Phi_{\text{ext}}/\Phi_0\!\!=\!\!0.3$ in Fig.\ \ref{bandmap}(a)]. As a result, the wavefunction of the fourth excited state slightly spreads out of the $m=1$ well. For even higher levels, the inter-band coupling becomes stronger and the classification into metaplasmon and persistent-current states ceases to apply.  

The situation of half-integer and integer $\Phi_\text{ext}/\Phi_0$ is special because of the additional parity symmetry of the potential $V(\varphi)$. For $\Phi_\text{ext}/\Phi_0\!= \!0.5$, the state $|\!-\!\!1_m,0_s\rangle$ becomes degenerate with $|0_m,0_s\rangle$. The ground and first excited states are hence the symmetric and antisymmetric superposition of 
$|\!-\!1_m,0_s\rangle$  and $|0_m,0_s\rangle$. For zero external flux, the ground state is $|0_m,0_s\rangle$, while the first and second excited states become the symmetric and antisymmetric superposition of $|\!-\!\!1_m,0_s\rangle$ and $|1_m,0_s\rangle$.

With the classification of states in hand, we now employ the analytic results from Sec.~\ref{analytical} to describe the qualitative behavior and magnitudes of the charge matrix elements.  Away from the degeneracies at integer and half-integer $\Phi_{\text{ext}}/\Phi_0$, the charge matrix element between the ground and the first excited state is approximated by the matrix element between two different persistent-current states, namely
\be
\langle 0_l| \mathsf{N} | 1_l \rangle \approx \langle 0_m,0_s| \mathsf{N}|\!-\!\!1_m,0_s\rangle.
\ee
Here, $l$ enumerates the fluxonium eigenstates in the order of their eigenenergies. The magnitude of this matrix element is relatively small because of the suppression enforced by the asymptotic selection rules for two persistent-current states [$\Delta s\!=\!\pm 1$ and $\Delta m=0$; see Eq.\ \eqref{pp2}]. Figure \ref{matrixtheory} shows that the charge matrix element between ground and first excited state, $\abs{\mathsf{N}_{01}}$, is indeed significantly smaller than the other elements (especially compared to $\abs{\mathsf{N}_{02}}$) over most of the flux range. 

We note this discrepancy in matrix element magnitudes could possibly lead to an interesting potential application: if coupling to the environment via charge dominates the qubit relaxation, then the lowest three fluxonium levels ($l=0,1,2$) could form a $\Lambda$-system over a wide flux range, with the  state $|1_l\rangle$ being a relatively long-lived metastable state, as noted previously in Ref.\ \onlinecite{Vlad-thesis}.  The origin of the $\Lambda$-configuration is intuitive from Fig.\ \ref{bandmap}(b): the states $|0_l\rangle$ and $|1_l\rangle$  are persistent-current states localized in different minima with only very small wavefunction overlap. The state $|2_l\rangle$, by contrast, is a metaplasmon state and has a large wavefunction overlap with both persistent-current states, resulting in relatively large matrix elements (and hence transition rates) between these states.  As a result, the state $|1_l\rangle$ may have a significantly longer life time than the state $|2_l\rangle$.

It is instructive to assess the deviation of exact results for the experimental parameters from the asymptotic prediction. For this comparison, we choose two eigenstates which, in a given flux range, can be approximately classified as a metaplasmon state and a persistent-current state, respectively. The approximate metaplasmon state we choose is $|0_\nu,1_s\rangle$.  Figure \ref{bandmap}(a) shows that, in the flux region $0$$<$$\Phi_\text{ext}/\Phi_0$$<$$0.1$, this metaplasmon state approximates the state $|3_l\rangle$. In the remaining flux region, this  metaplasmon state approximates the state $|2_l\rangle$.    The persistent-current state we choose is $|0_m, 0_s\rangle$, which approximates the ground state $|0_l\rangle$ away from $\Phi_\text{ext}/\Phi_0=0.5$.  We employ Eq.\ \eqref{pm} to calculate the asymptotic result for the matrix element $|\langle 0_m,0_s | \mathsf{N} | 0_\nu,1_s\rangle|$, where the input parameter $\epsilon_{1,1}/h=1.774 \  \text{GHz}$ is acquired from the half-width of the $s=1$ CPB band by diagonalizing $H'$.

The result obtained from Eq.\ \eqref{pm} is valid only sufficiently away from $\Phi_\text{ext} /\Phi_0$$=$$0.5$. There,
 the ground state $|0_l\rangle$ becomes a hybridization of the two states, $|\!-\!\!1_m,0_s\rangle$ and $|0_m, 0_s\rangle$, which are energy degenerate states in the absence of $\epsilon_{0,1}$. To account for this, we consider the $2\times2$ subspace containing both persistent-current states. The Hamiltonian in this subspace is
\begin{align}
\nonumber H_\text{sub} \approx 
\begin{pmatrix}
2\pi^2 E_L (\frac{\Phi_\text{ext}}{\Phi_0})^2 + \epsilon_{0,0} & \frac{1}{2}\epsilon_{0,1} \\
\frac{1}{2}\epsilon_{0,1} & 2\pi^2 E_L (\frac{\Phi_\text{ext}}{\Phi_0}-1)^2+ \epsilon_{0,0}
\end{pmatrix}
,
\end{align}
a truncated version of Eq.\ \eqref{Hs2}. Here, the input parameter $\epsilon_{0,1}/h=0.187 \ \text{GHz}$ is obtained from the half-width of the $s=0$ CPB band.   By diagonalizing this matrix, we obtain an improved approximation for the ground state $|0_l\rangle$, and for the asymptotic prediction of the matrix element $|\langle 0_l | \mathsf{N} | 0_\nu,1_s\rangle|$.  

The asymptotic prediction (dashed curve in Fig.\ \ref{matrixtheory}) is to be compared with the corresponding solid curves showing the numerically exact results -- in particular, the results for $\mathsf{N}_{03}$ and $N_{02}$ in the previously mentioned flux ranges.
Agreement is qualitative rather than quantitative, as expected. Note that, by accounting for hybridization, the  complete suppression of $\mathsf{N}_{02}$ at $\Phi_\text{ext}/\Phi_0$$=$$ 0.5$ enforced by parity symmetry is correctly predicted.
 Similarly, the asymptotic prediction for vanishing $\mathsf{N}_{01}$ agrees qualitatively with the significantly smaller values obtained numerically.

\vspace{0.01in}

\section{Conclusion\label{sec:conclusion}}
In summary, we have derived asymptotic expressions for the charge matrix elements of the fluxonium circuit in the parameter limit $E_J \gg E_C \gg E_L$, presented in Eqs. \eqref{pp2}--\eqref{pm}. Our derivation is based on the classification of fluxonium eigenstates into persistent-current and metaplasmon states \cite{koch_charging_2009}, and produces simple selection rules for the band indices $s$ and other quantum numbers which can be  intuitively understood from the localization properties of the different types of states.

   We employ our asymptotic predictions to interpret the numerically calculated matrix elements for the intermediate parameter regime realized in experiments \cite{manucharyan_fluxonium:_2009}. Even though quantitative agreement cannot be expected in this intermediate regime, 
we find good qualitative agreement and confirm that the asymptotic selection rules provide a useful predictor for different magnitudes of charge matrix elements.   Thus, our results can easily guide the choice of experimental parameters in order to reach the desired tunability of charge matrix elements in future fluxonium devices. The relatively large degree of tunability in fluxonium devices can be harnessed for influencing transition rates (possibly providing a route towards a $\Lambda$-system \cite{Vlad-thesis}), dispersive shifts \cite{guanyu_dispersive2012}, as well as the effective qubit-qubit interaction strength when coupling multiple fluxonium devices to a single microwave resonator mode \cite{majer_coupling_2007}.


\begin{acknowledgments}
We are indebted to Michael Devoret, Leonid Glazman and Vladimir Manucharyan for numerous insightful discussions.  Our research was supported by the NSF under Grant No. PHY-1055993.
\end{acknowledgments}

\end{document}